\documentclass[aps,nofootinbib,twocolumn,showpacs,preprintnumbers,amsmath,amssymb]{revtex4}

\usepackage{graphicx}% Include figure files
\usepackage{dcolumn}% Align table columns on decimal point
\usepackage{bm}% bold math

\newcommand{\ba}{\begin{eqnarray}}
\newcommand{\ea}{\end{eqnarray}}
\newcommand{\bc}{\begin{center}}
\newcommand{\ec}{\end{center}}

\newcommand{\ovl}{\overline} % Long Overline
\newcommand\chiM[2]{\chi_{{}_{#1}}^{\,#2}}
\newcommand\ovlchiM[2]{\ovl\chi_{{}_{#1}}^{\,#2}}

\newcommand\longit[1]{#1_{\!{}_\ell}}

\begin{document}

\title{Double heavy-quarkonium production from electron-positron annihilation
in the Bethe-Salpeter formalism}

\author{Xin-Heng Guo$^1$ }
\author{Hong-Wei Ke$^2$}
\author{Xue-Qian Li$^2$}
\author{Xing-Hua Wu$^3$}
 \affiliation{$^1$Institute of Low Energy Nuclear Physics,
Beijing Normal University, Beijing 100875, China,
\\$^2$Department of Physics, Nankai
University, Tianjin
 300071, China,\\
 $^3$College of Physics and Information Engineering, Henan Normal University, Xinxiang 453007, China}

\date{\today}

\begin{abstract}
In this work we evaluate the cross section of the process $e^+e^-\to
J/\psi\,\eta_c$ at energy $\sqrt{s}\approx 10.6$ GeV in the
Bethe-Salpeter formalism. To simplify our calculation, the heavy
quark limit is employed. Without taking the beyond-leading-order
contribution(s) into account, the cross section calculated in this
scenario is comparable with the experimental data. We also present
our prediction for the cross section of double bottomonium
production $e^+e^-\to \Upsilon(1S)\eta_b$ for the energy range of
$\sqrt{s}\approx (25\,\hbox{-}\,30)$ GeV which may be experimentally
tested, even though there is no facility of this range available at
present yet.
\end{abstract}


\maketitle

\section{introduction}

It is well known that there is a significant discrepancy between the
experimental measurements \cite{belle,babar} and the NRQCD
predictions \cite{NRQCD,e2} for the process $e^+e^-\to
J/\psi\,\eta_c$ at energy $\sqrt{s}\approx 10.6$ GeV. To reduce the
discrepancy many efforts have been made. For example, as discussed
in Ref.~\cite{e1}, corrections from pure electromagnetic
interactions are introduced into the non-relativistic QCD (NRQCD)
factorization formalism; the next-to-leading-order contributions of
strong interaction are taken into account in Refs.~\cite{e3,wangjx};
in Ref.~\cite{e7}, the authors take into account corrections to the
$J/\psi$ leptonic decay width and the scale dependence of the
leading-order prediction; etc.

Indeed, NRQCD should work well while dealing with the processes
where heavy flavors are involved, especially in this concerned
reaction where only heavy flavors exist. In the scenario of NRQCD an
expansion is carried out with respect to powers of the relative
three-velocity $v$ which reflects the relativistic effects. Even
though, in this reaction, the relativistic effects obviously are
smaller than in the processes dominated by light flavors, they still
may cause sizable contributions. Later works confirm this viewpoint.
One needs, therefore, to incorporate the relativistic corrections
into the theoretical calculations in an appropriate way. There are
various ways to take into account the relativistic corrections.
Ebert et al. carefully studied the corrections from relativistic
effects \cite{ebert} and indicated that with the relativistic
effects being properly handled, a prediction for the cross section
which is consistent with the present experimental measurement can be
obtained.

Moreover, the factorization scheme should work in this case, because
the production of $J/\psi+\eta_c$ from $e^+e^-$ via electromagnetic
interaction can be regarded as a reversed Drell-Yan process where
the factorization is proved. Thus one can factorize the hard and
soft processes and then a convolution integration over the two parts
results in the final amplitude. Therefore, one only needs to
consider the relativistic corrections appearing in the soft part,
i.e. in the hadronization process because the hard part is carried
out in the framework of quantum field theory which is completely in
the relativistic covariant form.

To understand the experimental results, many authors have proposed
various projects to improve the theoretical framework in addition
to NRQCD. For example, this problem was discussed with the method
of perturbative QCD (pQCD) . In Ref.~\cite{e5}, the corrections of
higher-twist wave functions were included in pQCD and the
light-front quark model.

Discussions given  in Refs.~\cite{e3,wangjx,e6,hfc-0702239} suggest
that to reduce the large discrepancy between the experimental
results and the theoretical predictions based on NRQCD for the
exclusive process with  the final state which is composed of two
charmonia, large next-to-leading-order (NLO) corrections may appear
(the `NLO' contribution is about 1.8 to 2.1 times of the
leading-order one). Including this large NLO contributions, their
results are close to the lower bound set by the Babar and Belle
collaborations for the double-charmonia production. The authors also
indicate that including the relativistic corrections can further
enhance the estimated value.

%Our work is along another line which is similar to the work of
%Ebert's, to investigate the relativistic effects on the production
%rate, in a different theoretical framework.

We would rather incline to believe that because of the validity of
the factorization, perturbative calculation is suitable and,
therefore, in our calculation we will ignore the contribution of the
next-to-leading order corrections. Then one should expect that a
larger correction may come from the soft part, i.e. the relativistic
effect may be significant.

As mentioned before, Ebert et al. considered such effects
\cite{ebert}. Alternatively, in this work, for properly
incorporating relativistic effects, we try to evaluate this
exclusive process in the Bethe-Salpeter (BS) formalism \cite{bs1}.
The BS equation is in principle established in the framework of
relativistic quantum field theory, therefore, it is supposed to
include all relativistic effects. Of course, to solve this equation
in practice, one needs to adopt some approximations such as the
instantaneous approximation where part of the relativistic effects
are lost. However, in many cases, such loss is not serious. Indeed,
the BS formalism is suitable for studying the binding systems
composed of two heavy charm and anti-charm quarks (or bottom
flavors). The transition amplitudes can be obtained, in a natural
way, to be an overlap integration of BS wave functions (see e.g.
Ref.~\cite{bs2}). In order to simplify the calculation, we will
further impose the heavy quark limit \cite{HQET1,HQET2,HQET}, i.e.
all the $1/m_Q$ corrections are neglected in the calculation in this
paper. Under this limit, our result shows that, at the leading order
of $\alpha_s$, the theoretical prediction is comparable with the
experimental result \cite{belle,babar}. We attribute this mainly to
the inclusion of  relativistic effects in our formalism. We will
come back to give more discussion on this point in the last section.

The remainder of this paper is organized as follows. In
Sec.~\ref{sect: bse}, we will study the BS equations for the vector
and pseudo-scalar quarkoniums. In Sec.~\ref{sect: pair-production},
we will calculate the cross section of the process $e^+e^-\to
J/\psi\, \eta_c$ in the BS formalism. In this section, we will also
give the prediction for the much smaller cross section of the
exclusive process with double bottomonium production,
$e^+e^-\to\Upsilon(1S) \eta_b$. Sec.~\ref{sect: conclusion} is
reserved for our conclusions and discussions.

\section{BS equations for heavy quarkoniums}
\label{sect: bse}

The BS wave function for a meson which is composed of a quark and an
anti-quark is defined as \ba \chiM{P}{}(x_1,x_2)_{\alpha\beta}
={\delta_{ij}\over \sqrt{3}}\, \langle 0|{\rm
T}\,\psi_\alpha^i(x_1)\ovl\psi_\beta^j(x_2)|P\rangle \,. \ea where
$P$ is the momentum of the meson, $\psi_\alpha^i(x_1)$ and
$\ovl\psi_\beta^j(x_2)$ are the field operators of the quark and
anti-quark, respectively, $\alpha, \beta$ are Dirac spinor indices,
and $i,j$ denote the color indices. In momentum space the BS
equation for the wave function $\chiM{P}{}(x_1,x_2)_{\alpha\beta}$
can be written as (see Refs. \cite{bs3,bs4} for example)
\begin{eqnarray}
\chiM{P}{}(p)&=&\frac{i}{p\!\!\!\slash_1-m_1+i\epsilon}
\int\frac{d^4k}{(2\pi)^4}V_P(p,k)\chiM{P}{}(k)
\nonumber\\ &&\times\,
\frac{i}{{\not\!p}_2-m_2+i\epsilon}\,,
\label{bs1}
\end{eqnarray}
where $p, k$ represent the relative momenta between the quark and
anti-quark, $m_1 (p_1)$ and $m_2 (p_2)$ are the masses (momenta) of
the quark and anti-quark, respectively, and $V_P(p,k)$ is the
kernel. The spinor indices are suppressed for simplicity. For the
heavy quarkonium $Q\ovl Q$ ($Q=c,b$) studied in this paper, we have
$m_1=m_2=m_Q$ and then $p_1=P/2+p$ and $p_2=-P/2+p$.

To simplify the calculation we will take the heavy quark limit
throughout this paper. In this limit, the propagators of the heavy
quark and heavy anti-quark ($S(p_1)$ and $S(p_2)$, respectively) can
be simplified in the following way \cite{heavy-baryon}: \ba
\frac{1}{{\not\!p}_1-m_Q+i\epsilon}
\to {(1+{\not\!v})/2 \over \longit{p}+E_0/2+i\epsilon} \,,\\
\frac{1}{{\not\!p}_2-m_Q+i\epsilon} \to {-(1-{\not\!v})/2 \over
\longit{p}-E_0/2-i\epsilon} \,, \ea where we have defined the
binding energy $E_0=M-2m_Q$, $v=P/M$ is the `four-velocity' of the
meson,  $M$ is the mass of the meson and $\longit{p}= v\cdot p$.
With this simplification, the BS equation (\ref{bs1}) becomes \ba
\chiM{P}{}(p)&=&{(1+{\not\!v})/2 \over \longit{p}+E_0/2+i\epsilon}
\int\frac{d^4k}{(2\pi)^4}V_P(p,k)\chiM{P}{}(k)\, \nonumber\\
&&\times\, {(1-{\not\!v})/2 \over \longit{p}-E_0/2-i\epsilon}\,.
\label{bs2} \ea From this expression, one can easily see that the BS
wave function $\chiM{P}{}$ satisfies
${\not\!v}\chiM{P}{}=\chiM{P}{}$ and
$\chiM{P}{}{\not\!v}=-\chiM{P}{}$. Then, in the heavy quark limit,
similar to the case for the diquark system studied in Ref.
\cite{bs5}, we have the following very simple parameterizations for
the BS wave functions of  vector and pseudoscalar $Q\ovl Q$ mesons
(which are denoted by the subscripts "$a$" and "$b$", respectively):
\ba \label{bsw-vector}
\chiM{P_a}{(s)}(p)&=&(1+{\not\!v}){\not\!\varepsilon}^{(s)}M_a
f_a(p)\,,\quad J^{PC}=1^{--}\,,
\\
\chiM{P_b}{}(p)&=&(1+{\not\!v})\gamma_5 M_b f_b(p)\,,\quad
J^{PC}=0^{-+}\,, \label{bsw-scalar} \ea where $\varepsilon^{(s)}$ is
the polarization vector of the vector quarkonium which is orthogonal
to the velocity, $\varepsilon^{(s)}\cdot v=0$, $f_a$ and $f_b$ are
scalar functions of the relative momentum $p$.

In this work, following the standard procedure for solving the BS
equation we impose the instantaneous approximation onto the kernel
as $V_P(p,k)=V({\bf p},{\bf k})$. Taking the concrete steps given in
Ref. \cite{bs3}, we can obtain this kernel by a Fourier
transformation of the Cornell potential which contains a linear
piece and a Coulomb-type piece, $-i V(r)=I\otimes I
V_s(r)+\gamma^\mu\otimes\gamma_\mu\, V_v(r)$, where $V_s(r)=\lambda
r+V_0$ and $V_v(r)=-\frac{4}{3}\frac{\alpha_s}{r}$. To avoid any
infrared divergence, a convergent factor $e^{-\beta r}$ is
introduced into the potential,
\begin{equation}
V_s(r)=\frac{\lambda}{\beta}(1-e^{-\beta
r})+V_0\,,\quad
V_v(r)=-\frac{4}{3}\frac{\alpha_s}{r}e^{-\beta r}.
\end{equation}
After the Fourier transformation, the kernel in momentum space
reads \cite{bs3}: $-i V({\bf q})=I\otimes I V_s({\bf
q})+\gamma^\mu\otimes\gamma_\mu\, V_v({\bf q})$, where \ba
V_s({\bf q})&=&-(\lambda/\beta+V_0)\delta^3({\bf q})+
\frac{\lambda}{\pi^2}\frac{1}{({\bf q}^2+\beta^2)^2}\,,
\nonumber\\
V_v({\bf q})&=&-\frac{2}{3\pi^2}\frac{\alpha_s({\bf q})}{({\bf
q}^2+\beta^2)}, \ea and the effective coupling constant is given by
$\alpha_s({\bf q})=4\pi/9\log{(a+\frac{{\bf
q}^2}{\Lambda^2_{QCD}})}$ with $a$ being a parameter which freezes
the running coupling constant at low energy.

With this kernel, the BS equation (\ref{bs2}) is written as \ba
\label{bsw-ins}
\chiM{P}{}(p)&=&{i\over(\longit{p}+E_0/2+i\epsilon)(\longit{p}-E_0/2-i\epsilon)}
{1+{\not\!v}\over 2} \nonumber\\&&\times\, \int\frac{d^4k}{(2\pi)^4}
\Big[ V_s({\bf q})\chiM{P}{}(k) + V_v({\bf
q})\gamma^\mu\,\chiM{P}{}(k)\,\gamma_\mu\Big] \nonumber\\&&\times\,
{1-{\not\!v}\over 2}\,, \ea where ${\bf q}={\bf p}-{\bf k}$.
Substituting the BS wave functions (\ref{bsw-vector}) and
(\ref{bsw-scalar}) into Eq. (\ref{bsw-ins}) we get the following
component equations for vector and pseudo-scalar quarkonia: \ba
f_{a(b)}(p)&=&{i \over
(\longit{p}+E_0/2+i\epsilon)(\longit{p}-E_0/2-i\epsilon)}
\nonumber\\&&\times\, \int\frac{d^4k}{(2\pi)^4} (V_s-V_v)({\bf q})\,
f_{a(b)}(k)\,. \label{bsw-component-4d} \ea For later convenience,
here we also write out the BS equation for the conjugate wave
function, \ba \ovlchiM{P}{}(p)
&=&{i\over(\longit{p}+E_0/2+i\epsilon)(\longit{p}-E_0/2-i\epsilon)}
{1-{\not\!v}\over 2} \nonumber\\&&\times\, \int\frac{d^4k}{(2\pi)^4}
\Big[ V_s({\bf q})\ovlchiM{P}{}(k) + V_v({\bf
q})\gamma^\mu\,\ovlchiM{P}{}(k)\,\gamma_\mu\Big]
\nonumber\\&&\times\, {1+{\not\!v}\over 2}\,. \ea

To solve the BS equation (\ref{bsw-component-4d}), for convenience,
we can choose a coordinate frame in which the binding system is
static. The BS wave functions in this frame are given in Eqs.
(\ref{bsw-vector}) (\ref{bsw-scalar}), where $f_a(p)$ and $f_b(p)$
are Lorentz scalar functions. Then the longitudinal component
$\longit{p}=p^0$. After carrying out the integrations over $p^0$ and
$k^0$ on both sides of the BS equation along some proper contour,
Eq. (\ref{bsw-component-4d}) becomes
     \footnote{
       In pratical calculation, to obtain the reasonable BS wave
       functions, we will replace $E_0$ by $M_{a(b)}-2\sqrt{{\bf k}^2+m_Q^2}$
       in the following BS equation. This replacement is equivalent to
       regaining  some part of  $1/m_Q$ effects in the calculation.
     }
\ba \widetilde f_{a(b)}({\bf p})={1\over E_0}\int {d^3{\bf k}\over
(2\pi)^3} (V_v-V_s)({\bf q})\, \widetilde f_{a(b)}({\bf k})\,,
\label{bsw-component} \ea where  we have defined the instantaneous
wave functions by $\widetilde f_{a(b)}({\bf p})=\int {dp^0\over
2\pi} f_{a(b)}(p)$.

\bigskip\noindent
{\it Normalization of the BS wave functions.}
\bigskip

\noindent The normalization condition of the BS wave function
$\chiM{P}{}$ for a vector meson can be written as \ba &&i\int
{d^4p\,d^4p'\over (2\pi)^8}\ovlchiM{P_a}{(s)}(p)
\left[{\partial\over\partial P_a^0}I_{P_a}(p,p')\right]
\chiM{P_a}{(s')}(p') \nonumber\\&& =2P_a^0\delta_{ss'}\,, \ea where
$I_P(p,p')=-(2\pi)^4\delta^4(p-p')\,S^{-1}(p_1)S^{-1}(p_2)$\,, $s$
and $s'$ are the spin indices of the vector meson, and
$P_a^0=\sqrt{{\bf P}_a^2+M_a^2}$ is the energy of the bound state.
Multiplying by $\delta_{ss'}$ on both sides and summing over $s$ and
$s'$, one has the following normalization equation (in the static
frame of the meson): \ba {8 \, v_a^0\over E_0^2}\int {d^3{\bf
p}\over (2\pi)^3}\, (\widetilde f_{a1}-\widetilde
f_{a2})^2=2P_a^0\,, \label{vector-normalization} \ea for the vector
meson. $\widetilde f_{a1}$ and $\widetilde f_{a2}$ are defined as
follows, \ba \widetilde f_{a1}({\bf p})=M_a\int{d^3{\bf k}\over
(2\pi)^3}\, V_s({\bf p}-{\bf k})\widetilde f_a({\bf k})\,,
\\
\widetilde f_{a2}({\bf p})=M_a\int{d^3{\bf k}\over (2\pi)^3}\,
V_v({\bf p}-{\bf k})\widetilde f_a({\bf k})\,. \ea From the BS
equation (\ref{bsw-component}), $\widetilde f_{a1}-\widetilde
f_{a2}=- E_0 M_a\widetilde f_a$, one can see that the normalization
equation (\ref{vector-normalization}) can be written as \ba 4 \,
M_a\int {d^3{\bf p}\over (2\pi)^3}\, \widetilde f_a({\bf p})^2=1\,.
\label{vector-normalization-simplified} \ea The normalization
equation for the pseudo-scalar meson has completely the same form as
Eq. (\ref{vector-normalization-simplified}) (with the subscript
"$a$" replaced by "$b$").

\section{Double quarkonium production from $e^+e^-$ annihilation}
\label{sect: pair-production}

\begin{figure}[htb]
\centering
\includegraphics*[scale=0.68]{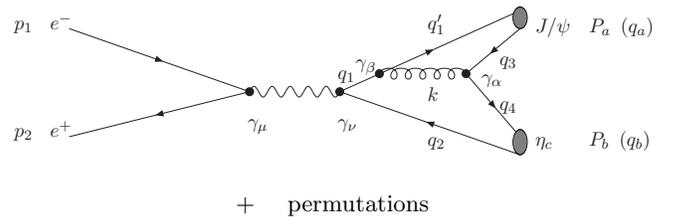} \\[10pt]
+ \quad permutations
\caption{\label{figure-1} Feynman diagrams for double charmonium production
from electron-positron annihilation.}
\end{figure}

Now we turn to discuss the exclusive process in the
electron-positron collisions with the final state of two heavy
quarkonia. The relevant Feynman diagrams for the process
$e^+e^-\to J/\psi\,\eta_c $ are depicted in Fig.~\ref{figure-1}.
Since the s-channel gluon turns into $c\bar{c}$ in this process,
hence the gluon is hard and the energy scale for this process is
large. Therefore, the perturbative calculation in QCD to lowest
order of $\alpha_s$ expansion is expected to be sufficient. We will
not consider the contributions from diagrams of higher-order in
$\alpha_s$. Furthermore, we do not consider any higher order
corrections from the pure electromagnetic interactions, which should
be small and were discussed in Ref.~\cite{e1}.

One of the amplitudes in Fig.\ref{figure-1} can be written as
%\begin{widetext}
\begin{eqnarray}
{\cal M}_1&=&C\,\frac{g^{\mu\nu}e\, e_Q^{}g_s^2}{s}\,\bar
v(p_2)\gamma_\mu u(p_1) \int {d^4q_ad^4q_b\over (2\pi)^8}\,
\nonumber\\ &&\times {\rm
Tr}\,\Big[\ovlchiM{P_a}{}\gamma_\beta\,{1\over {\not\!q}_1-m_c}\,
\gamma_\nu
\,\ovlchiM{P_b}{}\gamma_\alpha\Big]\,\frac{g^{\alpha\beta}}{k^2} .
\label{amplitude-1}
\end{eqnarray}
%\end{widetext}
where $C=4/3$ is the color factor, $\sqrt{s}$ is the total energy
in the center-of-mass frame, and $e^{}_Q$ is the electric charge
of the quark $Q$. The conjugation of the BS wave function is
defined by $\ovlchiM{P}{}\equiv\gamma^0(\chiM{P}{})^\dag\gamma^0$.
The momenta of the quark and anti-quark within the final state are
$q'_1=\frac{1}{2}P_a+q_a$\,, $q_3=\frac{1}{2}P_a-q_a$\,,
$q_4=\frac{1}{2}P_b+q_b$\,, $q_2=\frac{1}{2}P_b-q_b$\,, and the
momenta in the gluon and the quark propagators are given by
\begin{eqnarray}
&&k=q_3+q_4=\frac{1}{2}(P_a+P_b)-q_a+q_b\,,\\
&&q_1=q'_1+k=P_a+\frac{1}{2}P_b+q_b\,,
\end{eqnarray}
respectively. Since an integration is needed to obtain the amplitude
(\ref{amplitude-1}) and since the propagators of the quark and the
gluon depend on the relative momenta $q_a$ and $q_b$ one can expect
that the calculation is very complicated. To simplify the
calculation, we assume that the propagators of the quark and the
gluon are independent of relative momenta $q_a$ and $q_b$ (see, e.g.
Ref. \cite{bs5}). This simplification is appropriate since the
masses of heavy quarks are large compared with the relative momenta,
which are of order $\alpha_s m_Q$. Then the momenta $q_1$ and $k$ of
the propagators are large compared with the relative momenta $q_a$
and $q_b$. One may expect that, in the heavy quark limit, the
calculation without taking into account the relative momenta should
be exact.
   \footnote{
     The energy scale $\mu$ is of the same order of $m_Q$ and then, in the
     heavy quark limit, $\alpha_s(\mu)\sim 1/\log(m_Q/\Lambda_{\rm QCD})\to 0$.
   }

With the above approximation in mind, the momenta $k$ and $q_1$ can
be written as: $ k\approx (P_a+P_b)/2$, $q_1\approx P_a+P_b/2$ which
lead to $k^2\approx s/4$, $q_1^2\approx s/2+M_a^2/2-M_b^2/4\approx
s/2+m_c^2$. Furthermore, we will make use of the approximation
$M_{J/\psi}\approx M_{\eta_c}\approx 2m_c$ in the calculation
throughout this paper. Then the amplitude (\ref{amplitude-1}) can be
written as
%\begin{widetext}
\ba {\cal M}_1&=&\frac{2^6}{3^2s^3}g_s^2e^2 \, \bar v(p_2)\gamma_\mu
u(p_1)
\int\frac{d^4q_a}{(2\pi)^4}\frac{d^4q_b}{(2\pi)^4}\, \nonumber\\
&&\times {\rm
Tr}\,\big[\ovlchiM{P_a}{(s)}(q_a)\gamma_\alpha\,({\not\!q}_1+m_c)
\gamma^\mu \ovlchiM{P_b}{}(q_b)\gamma^\alpha\big].\quad \ea
%\end{widetext}
From the component expressions of the BS wave functions
(\ref{bsw-vector}) and (\ref{bsw-scalar}), we can see
$\ovlchiM{P_a}{(s)}={\not\!\varepsilon}^{(s)}(1+{\not\!v_a})M_a
f_a$ and $\ovlchiM{P_b}{}=\gamma_5(1+{\not\!v_b})M_b f_b$.
    \footnote{
      We have rotated the phase of the wave function to make $f_a$ and $f_b$
      be real.
    }
Substituting the BS wave functions into the trace in the above
amplitude, one achieves \ba {\cal M}_1 &=&-\,i\,{2^{14}\pi^2\over
3^2}\,{\alpha_{\rm s}\alpha_{\rm em}\over s^3}\,
m_c\,\epsilon_{\mu\nu\rho\sigma}\,\varepsilon^{(s)\nu} P_a^\rho
P_b^\sigma \nonumber\\&&\times\, \bar v(p_2)\gamma_\mu
u(p_1)\psi_a\psi_b\,, \ea where $\alpha_s=g_s^2/4\pi$,
$\alpha_{\rm em}=e^2/4\pi$, $\psi_a$ and $\psi_b$ are two numbers
defined by the integrations over $f_a$ and $f_b$, respectively,
\ba \psi_{a(b)}=\int\frac{d^4q}{(2\pi)^4}\,f_{a(b)}(q)
=\int\frac{d^3{\bf q}}{(2\pi)^3}\,\widetilde f_{a(b)}({\bf q})\,.
\ea The total amplitude for the process $e^+ e^-\to
J/\psi\,\eta_c$ can be obtained by summing over all the amplitudes
of the diagrams shown in Fig.~\ref{figure-1}.

The unpolarized total cross section (see e.g. Ref. \cite{PDG06}) is
obtained by summing over various $J/\psi$ spin-states and averaging
over those of the initial state $e^+ e^-$, \ba \sigma={1\over
32\pi}\,{\sqrt{s-16m^2_c}\over s^{3/2}} \int {1\over 4}\sum_{\rm
spins}|\mathcal{M}_{\rm total}|^2\,d\cos\theta \,, \ea where the
masses of the electron and positron are ignored in the calculation.
The explicit expression for the total amplitude $|\mathcal{M}_{\rm
total}|^2$, which is the sum of all diagrams shown in Fig.
\ref{figure-1}, is written as \ba \label{s2} {1\over 4}\sum_{\rm
spins}|\mathcal{M}_{\rm total}|^2 &=&{2^{30}\pi^4\over
3^4}\,{\alpha_{\rm em}^2\alpha_s^2\over s^5}\, m_c^2(-32 m_c^4  +
t^2 + u^2) \nonumber\\&&\times\, \psi_a^2 \psi_b^2 \,, \ea where
$t=(p_1-P_a)^2$ and $u=(p_1-P_b)^2$ are the Mandelstam's variables.

\bigskip\noindent
{\it Numerical results.}
\bigskip

\begin{figure}[htb]
\centering
\includegraphics*[scale=0.8]{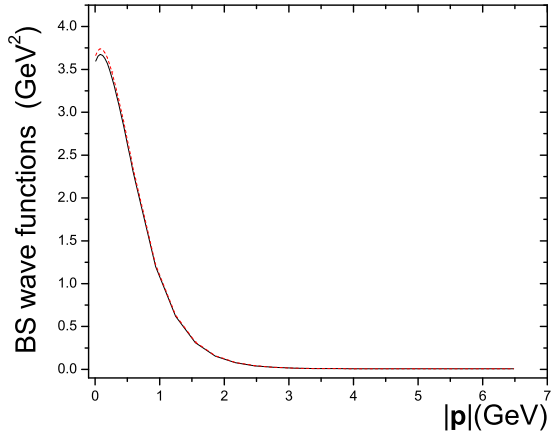}
\caption{\label{cc-wavefunction}
  BS wave functions for $J/\psi$ and $\eta_c$ in the heavy quark limit, $\widetilde f_a$ (solid line) and
$\widetilde f_b$ (dashed line), respectively. }
\end{figure}

\noindent The parameters in the calculation will be taken to be
\cite{bs2,bs3}: $a = 2.7183$, $\beta= 0.06$ GeV, $\alpha_{\rm
em}\approx 1/137$\,, $\alpha_s=0.26$, $\lambda$= 0.2 GeV$^2$,
$\Lambda_{\rm QCD}$ = 0.26 GeV, $m_c=1.7753$ GeV. In the interaction
kernel $V_0=0.415$ GeV for $J/\psi$ and $V_0=0.525$ GeV for
$\eta_c$. With these parameters, we can solve the BS equations
numerically and the wave functions $\widetilde f_a$ and $\widetilde
f_b$ are plotted in Fig.\ref{cc-wavefunction}.

Then we have \ba \psi_a=0.1020 \hbox{ GeV}\,,\qquad \psi_b=0.1037
\hbox{ GeV}. \ea Consequently, the total cross section is obtained
as \ba \sigma(e^+e^-\to J/\psi\,\eta_c)=22.3 \hbox{ fb} \ea

If we vary $m_c$ and $\alpha_s$  within $10\%$ we can get an
errors $\pm1.6 fb$.

\bigskip

\noindent The above analysis for the exclusive process $e^+e^-\to
J/\psi\,\eta_c$ can be applied, with only a little modifications, to
the exclusive process $e^+e^-\to\Upsilon(1S)\,\eta_b$.
$\eta_b(9434)$ is the lowest-lying pseudo-scalar $b\bar b$ state
(for discussions about $\eta_b(9434)$, see e.g. Ref. \cite{eta-b}).
The mass of the $b$ quark is $m_b=5.224$ GeV \cite{bs3}. In the
interaction kernel $V_0=0.62$ GeV for $\Upsilon(1S)$ and $V_0=0.64$
GeV for $\eta_b$. With these parameters, we can solve numerically
the wave functions $\widetilde f_a$ and $\widetilde f_b$, which lead
to
%Fig.\ref{bb-wavefunction}.
\ba \psi_a=0.1123 \hbox{ GeV}\,,\qquad  \psi_b=0.1124 \hbox{ GeV}.
\ea Then the cross section is predicted to be \ba \sigma(e^+e^-\to
\Upsilon(1S)\,\eta_b)=(0.16\,\hbox{-}\,0.06) \hbox{ fb} \ea for
the range of the total energy $\sqrt{s}=(25\,\hbox{-}\,30)$ GeV.
%
%
\iffalse
\begin{figure}[htb]
\centering
\includegraphics*[scale=0.35]{bb-wavefunction.eps}
\caption{\label{bb-wavefunction}
  BS wave functions $\widetilde f_a$ and $\widetilde f_b$ for
  $\Upsilon(1S)$ and $\eta_b$, respectively, in the heavy quark limit. }
\end{figure}
\fi

\section{Conclusions and discussions}
\label{sect: conclusion}

In this paper, we study the exclusive  processes of $e^+e^-$
annihilating into two quarkonia in terms of the BS formalism. We
find that, in the heavy quark limit, the cross section is
$\sigma[e^+ e^-\to J/\psi\,\eta_c]=22.3$ fb, which is compatible
with the Babar's data, $\sigma[e^+e^- \to J/\psi\,\eta_c]\geq
17.6\pm 2.8\pm 2.1$ fb \cite{babar}, and the Belle's data,
$\sigma[e^+e^- \to J/\psi\,\eta_c]\geq 25.6 \pm 2.8 \pm 3.4$ fb
\cite{belle}. Because the BS formalism is established based on the
relativistic quantum field theory, one has a strong reason to
believe that some (perhaps not all) relativistic effects are
automatically included in the calculations. The missing part may
come from the instantaneous approximation which is necessary for
solving the BS equation. Since the approximation is proved to be
reasonable in theoretical calculations for other similar
processes, we may be convinced that the missing part is not
significant. Thus we expect that the non-leading-order
contributions, from extra $1/m_Q$ which is indeed a relativistic
effect, and $\alpha_s$ corrections, should be small. Our result is
different from those given in
Refs.~\cite{e3,wangjx,e6,hfc-0702239} in the NRQCD framework,
where the leading-order contribution is too small to be comparable
with the experimental data. In order to reduce the discrepancy
between the experimental data and the theoretical predictions
based on NRQCD, large non-leading-order contributions (including
$\alpha_s$ or/and relativistic correction(s)) are required in the
NRQCD framework and the value of the total non-leading-order
correction is nearly twice of the leading-order one.

We also calculate the cross section for the exclusive process with
two bottomonia as the final state,
$\sigma[e^+e^-\to\Upsilon(1S)\eta_b]=(0.16\,\hbox{-}\,0.06)$ fb
corresponding to the range of the total energy $\sqrt{s} =
(25\,\hbox{-}\,30)$ GeV. Since $m_b\gg m_c$, the non-leading-order
contributions, including effects of higher orders in $1/m_b$ and
$\alpha_s(2m_b)$ expansions, should be much smaller than those in
the charmonium case. Therefore, we expect that the calculation in
the bottomonium case is much more precise than that in the
charmonium case. Even though such processes cannot be measured at
present due to the constraint of the available energy range at the
B-factories, the planned ILC will be a powerful facility to testify
this result.

Due to the obvious advantage of the BS formalism for dealing with
the processes where heavy flavors are involved, we may hope that
the obtained results are close to reality.

%One notices the fact that the obtained cross section is still
%smaller than the lower bounds set by the experimental measurement
%even though there are no definite data, so it may imply that there
%exist some mechanisms which are not included in our discussions and
%we will work on this issue in our coming work.

It will be interesting to study the non-leading-order contributions
in the formalism used in this paper and check whether our
expectation about the non-leading-order contributions is right. This
task is beyond the scope of this paper and will be discussed
elsewhere.

\begin{acknowledgments}

This work was supported in part by National Natural Science
Foundation of China (Project Number: 10475042, 10675022), the Key
Project of Chinese Ministry of Education (Project Number: 106024),
Ph.D. Program Foundation of Ministry of Education of China and the
China Postdoctoral Science foundation (No. 20020055016) and the
Special Grants from Beijing Normal University.

\end{acknowledgments}

\end{document}